\documentclass{emulateapj}
\usepackage[usenames]{color}
\usepackage{graphicx}
\usepackage{amssymb}
\usepackage{amsmath}
\usepackage{times}
\newcommand{\NEW}[1]{{#1}}

\newcommand{\msun}{M_\odot}

\newcommand{\mbh}{M_\bullet}
\newcommand{\mc}{M_\mathrm{c}}
\newcommand{\md}{M_\mathrm{d}}
\newcommand{\unit}[1]{#1_\mathrm{u}}
\newcommand{\dist}[1]{n_{#1}}

\newcommand{\irms}{i_\mathtt{rms}}

\newcommand{\nbody}{$N$-body}
\newcommand{\degm}{^\circ}
\newcommand{\rd}{R_\mathrm{d}}
\newcommand{\pc}{\mathrm{pc}}
\newcommand{\kms}{\mathrm{km\,s^{-1}}}
\newcommand{\model}[1]{%
\ifnum#1=352{\sf{}M1}\else
\ifnum#1=347{\sf{}M2}\else
\ifnum#1=350{\sf{}M3}\else
\ifnum#1=345{\sf{}M4}\else
\ifnum#1=344{\sf{}M5}\else
\ifnum#1=341{\sf{}M6}\else
\ifnum#1=342{\sf{}M7}\else
\ifnum#1=321{\sf{}M8}\else
\ifnum#1=331{\sf{}M9}\else
{\tt ???}\fi\fi\fi\fi\fi\fi\fi\fi\fi
}
\shorttitle{Kozai-Lidov dynamics in eccentric stellar discs}
\shortauthors{Haas \& \v{S}ubr}
\begin{document}
\title{Rich Kozai-Lidov dynamics in an initially thin and eccentric stellar disc around\\a supermassive black hole}
\author{Jaroslav Haas and Ladislav \v{S}ubr}
\affil{Charles University in Prague, Faculty of Mathematics and Physics, Astronomical Institute, V Hole\v{s}ovi\v{c}k\'{a}ch 2, Praha, CZ-18000, Czech Republic}
\thanks{E-mail: haas@sirrah.troja.mff.cuni.cz; subr@sirrah.troja.mff.cuni.cz}
\begin{abstract}
There is growing evidence of star formation in the vicinity of supermassive black
holes (SMBH) in galactic nuclei. A viable scenario for this process assumes infall of
a massive gas cloud towards the SMBH and subsequent formation of a dense accretion disc
which gives birth to the young stars. Numerical hydrodynamical models indicate that
this star formation process is rather fast and it precedes full circularization of
the accretion flow, i.e. the new stars are born on elliptic orbits.
By means of direct numerical {\nbody} modeling, we show in this paper that the non-zero
eccentricity of the stellar discs around the SMBH leads to an onset of various types of the
Kozai-Lidov oscillations of a non-negligible subset of individual orbits in the disc,
showing a remarkable robustness of this classical mechanism.
Among others, we demonstrate that under certain
circumstances, presence of an additional spherical cluster (which is generally known
to damp Kozai-Lidov oscillations) may trigger such oscillations due to affecting the
internal flow of the angular momentum through the disc.
We conclude that the Kozai-Lidov oscillations are capable to substantially modify the
initial structure
of the disc (its thickness and distribution of eccentricities, in particular).
\end{abstract}
\keywords{Galaxy: nucleus --- stars: kinematics and dynamics --- celestial mechanics}
\section{Introduction}
The discovery of a young stellar disc in the central parsec of the Milky Way
\citep{Levin03} and observations of a similar structure in our neighboring galaxy
M31 in Andromeda \citep{Bender05} suggested the possibility that flattened stellar
structures might represent a generic component of galactic nuclei. Since then, many
aspects of the dynamical evolution of such systems have been investigated
mostly in the context of the disc in the center of the Milky Way
which is by far the best observationally accessible because of its proximity.

Diffusive processes due to two-body relaxation in an isolated disc of stars orbiting
a supermassive black hole (SMBH) were analyzed, e.g., in \citet{Alexander07} or
\citet{Cuadra08}\footnote{On the scale of planetary systems, this topic
was studied even earlier \citep[see][and references therein]{Stewart00}.}.
These works described gradual thickening of the disc and growth of eccentricity of
the individual stellar orbits.
Two-body relaxation among the stars of the disc was also the key elementary process
beyond the evolution of the radial structure of the disc studied by \citet{Subr14}.
\NEW{Resonant relaxation among the stellar orbits forming the disc was discussed by
\cite{Tremaine98}.} Later on, this process between the disc and an embedding spherical
star cluster was addressed by \citet{Kocsis11} who found that this process
may lead to a significant warp of the disc. \NEW{Angular momentum transfer through an
eccentric stellar disc embedded in a spherical cluster was investigated by
\cite{Madigan09}.}
Other works \citep[e.g.][]{Nayakshin06,Subr09,Gualandris12} considered various
additional external sources of a perturbative gravitational potential
in order to explain the properties of the observed structures of young stars
in the Galactic Center through secular dynamics.

Some of the above introduced works consider the stellar disc to be formed by initially
eccentric orbits. According to works of \citet{Bonnell08}, \citet{Wardle08},
\citet{Hobbs09} or \citet{Mapelli12}, such a setting is a probable outcome of the
currently most widely accepted formation scenario for the young stellar disc in the
center of the Milky Way via partial capture of an infalling massive gaseous cloud
by the central SMBH. \NEW{Models of eccentric disc of young stars were also considered
to explain the observed properties of the nucleus of M31
\citep[e.g.][]{Tremaine95,Peiris03}.} Presence of initially
eccentric orbits in the disc broadens the variety of possible modes of angular
momentum exchange throughout the disc. In our paper, we focus on the numerous effects
of the Kozai-Lidov dynamics in the potential of the disc itself and of the embedding
spherical potential.

\section{Kozai-Lidov dynamics}
\label{sec:KL}
In contrast to the classical, purely self-gravitating star clusters, the
motions of stars in central mass dominated systems are highly regular. This leads
to their mutual interaction on time-scales greatly exceeding the typical
orbital periods around the center. In order to describe the
resulting slow (secular) evolution of their orbits, it is very useful to think of
the whole system as averaged over one orbital revolution. This
approach (commonly called the averaging technique) has been developed in the
perturbation theory of the celestial mechanics and represents a well justified and
widely used tool for such a problem which relies on the classical Hamiltonian formalism
\citep[see, e.g.,][]{Morbidelli02,Bertotti03}. In the sense of the averaging technique,
the individual orbits can be also thought as wires exerting torques on one another
which directly affect their angular momenta; note, however, that energy (semi-major axis)
is conserved within the averaging approximation.
\begin{figure*}
\begin{center}
\includegraphics[width=\textwidth]{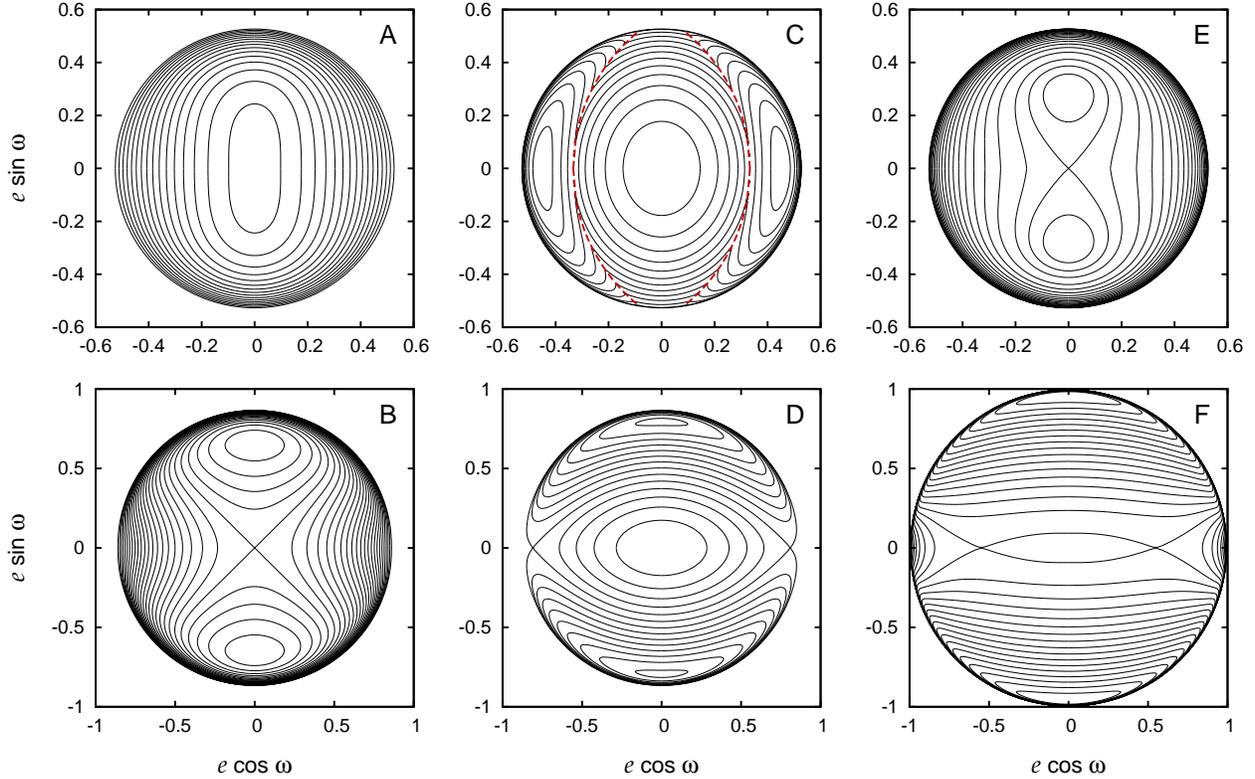}
\caption{\label{fig:KL-isocontours}
Isocontours of the averaged perturbative potentials in the $e$-$\omega$ space for
several different settings.
The perturbing potential is due to an infinitesimally thin ring of radius $\rd$ in panels
{\sf A} through {\sf C}; in panel {\sf D} the perturbing potential is a superposition
of potentials of a ring and a spherical cluster with radial density profile
$\varrho(r) \propto r^{-7/4}$ and mass $\mc = 1.5\md$ within the radius $\rd$. In panels
{\sf E} and {\sf F} the source of the perturbing potential is a razor-thin disc of constant
surface density and outer radius $\rd$. Specific values of orbit semi-major axis, $a$,
and the Kozai integral, $c$, are:
{\sf A}: $a=0.48\,\rd,\; c=0.85$;
{\sf B}: $a=0.48\,\rd,\; c=0.5$;
{\sf C}: $a=1.50\,\rd,\; c=0.85$;
{\sf D}: $a=0.48\,\rd,\; c=0.5$;
{\sf E}: $a=0.48\,\rd,\; c=0.85$;
{\sf F}: $a=0.48\,\rd,\; c=0.1$.
Dashed line in panel {\sf C} corresponds to orbits which intersect the perturbing ring.
Note different ranges of the boxes in the upper and lower panels which reflect
different maxima of eccentricity for different values of $c$.}
\end{center}
\end{figure*}

The averaging technique can be straightforwardly used to determine the secular evolution of a Keplerian orbit under the perturbative influence of a flattened potential.
Such a potential can represent either some rather continuous matter distribution (e.g. gaseous structures) or, in the context of the averaging technique, the averaged potential of a body or a disc of bodies.
The common feature of the flattened disturbing potentials is that they decrease the degree of symmetry of the unperturbed Keplerian potential. As a result, the vector of angular momentum of the Keplerian
orbit is no longer an integral of motion, permitting various ways of its evolution, either
periodic or chaotic.

A systematic study of the secular evolution of a Keplerian orbit due to presence of
a perturbing flattened potential began with the works of \citet{Kozai62} and
\citet{Lidov62} who studied the simplest form of this problem, the secular evolution
of the hierarchical three-body problem, i.e. a system in which two bodies form
a well-defined binary and the third body orbits around this binary staying well
separated from it. In these pioneering works, one of the components of the inner
binary has been considered massless and the orbit of the distant third body circular,
making its (averaged) perturbing potential axially symmetric and keeping the
projection of the angular momentum of the test orbit onto the symmetry axis an
integral of motion. Due to conservation of the orbital semi-major axis, $a$, of 
the test orbit (within the averaged approximation), the conserved
projection of the angular momentum may be replaced by the so-called Kozai integral,
$c\equiv\sqrt{1-e^2}\cos{}i$, where $e$ and $i$ are
eccentricity and inclination of the test orbit measured from the symmetry axis,
respectively. Hence, the Kozai integral enables us to eliminate $i$ (say) from
further considerations. Since the nodal longitude, $\Omega$, of the orbit
does not affect its evolution due to symmetry of the potential, the
remaining problem is to solve coupled equations for secular evolution of eccentricity
and argument of pericenter, $\omega$.

Within the averaging approximation, the perturbing potential averaged over one
Keplerian revolution around the center is another (third) integral of motion.
When the concern is about the trajectory in the phase space rather
than about temporal evolution of the orbital elements, it is possible to by-pass
the problem of solving differential equations by directly investigating
isocontours of the averaged perturbing potential in the $e$-$\omega$ space which represent the allowed trajectories.
The exemplary diagrams that reveal qualitatively different configurations are shown
in Figure~\ref{fig:KL-isocontours}.

We start with the classical setup, when the perturbation is due to a distant third
body on a circular orbit of radius $R_\mathrm{d}$ and the multipole expansion of
its averaged potential in ratio $a/R_\mathrm{d}$ is truncated after the quadrupole
term \citep[for details, see, e.g.,][]{Kozai62}.
Panel A corresponds to initial conditions for which the test
orbit does not undergo any significant oscillations of eccentricity and $\omega$ rotates
in the full interval $\langle0,2\pi\rangle$, following the simple oval-shaped isocontours.
Such a case occurs whenever the orbit fulfills the criterion $c>\sqrt{3/5}$ (the so-called
Kozai limit which corresponds to $i\lesssim 39.2^\circ$ for the circular orbit).
The topology of the isocontours changes dramatically for $c\leq\sqrt{3/5}$ (panel B). Two new stationary points at $\omega=\pi/2$ and $3\pi/2$ appear and a separatrix curve emerging from the origin
separates two regions in which $\omega$ librates in a limited interval from the outer region in which it still circulates.
The two stationary points correspond to solutions during which two angles,
$\varpi \equiv \Omega+\omega$ and $\Omega$ are in 1:1 resonance, leaving $\omega$ constant
during the evolution. For this reason, this type of topology is commonly referred to as resonant and we will adopt this nomenclature further on.
The most important feature of the resonant topology are high-amplitude oscillations
of eccentricity, forcing the orbit to a very high eccentricity state no matter how small its
initial eccentricity was. In the honor of their discoverers, these oscillations are
commonly called Kozai-Lidov oscillations and the whole phenomenon Kozai-Lidov resonance
(cycles; mechanism). Due to existence of the Kozai integral $c$, the test orbit is thus oscillating between
two extremes: (i) low eccentric and highly inclined with respect to the plane of symmetry of
the potential (perturbing orbit), and (ii) highly eccentric and nearly coplanar. Another
immediate consequence of the Kozai integral is that the inclination of the
test orbit may never cross value $i=\pi/2$, in other words, the orbital motion of the
test particle around the central mass remains either prograde
or retrograde with respect to the motion of the third body, never changing this sense.

Over the many decades since the original discovery, many other variants and
generalizations of the classical Kozai-Lidov problem have been studied extensively.
Among them, let us briefly comment on those which we found to be relevant for the
numerical models presented below. It was shown by \citet{Bailey92} and further
systematically studied by \citet{Thomas96} or \citet{Gallardo12}, that the dynamics
in the case when the perturbing body is inside the orbit of the test particle
is very much different in comparison with the classical setting (panel C in
Figure~\ref{fig:KL-isocontours}). Most notably, the initially circular orbit is not
an unstable solution of the problem, i.e., in order to undergo significant oscillations,
the test orbit must possess (or reach due to some other processes) a certain non-zero
eccentricity. Furthermore, in contrast to the classical setting, the stationary points
in the resonant topology are located at $\omega=0$ and $\pi$ and, for lower values
of the Kozai-Lidov integral, even more libration regions may appear. Let us also note
that if the test orbit undergoes significant oscillations of its eccentricity, it may
cross the orbit of the perturbing body at which point the assumptions of the
approximation are no longer valid. Therefore, the affected isocontours of
the perturbing potential may not correctly represent the real dynamics of the test
orbit. The dashed line in panel C connects the points for which this occurs. 

It was further shown that the Kozai-Lidov oscillations can be inhibited by embedding
the three-body system into some (sufficiently strong) spherical gravitational potential
\citep[e.g.,][]{Ivanov05,Subr07,Haas11}.
The reason for this is the precession of the argument of pericenter $\omega$ of the
test orbit caused by the spherical potential which disturbs the Kozai-Lidov resonance.
In terms of isocontours of the perturbing potential, superposition
of the potential of the ring with the spherical component leads to formation
of an inner rotational region which we illustrate in panel D
in Figure~\ref{fig:KL-isocontours}. We can see that, similarly to the case of the
interior perturber, it is necessary for the test orbit to possess some non-zero
eccentricity in order to undergo significant Kozai-Lidov oscillations. Moreover,
increasing strength of the spherical perturbing potential leads to a decrease of the
limiting value of the Kozai-Lidov integral, $c$, below which the resonance occurs,
i.e. it decreases the volume of the resonant part of the phase space.

In our numerical models, the source of the perturbing potential is represented by
a moreorless thin disc rather than an infinitesimally thin ring. As an example,
we plot isocontours for a razor-thin axially symmetric disc of a
constant surface density ranging from $r=0$ to $r=R_\mathrm{d}$ in panels
E and F of Figure~\ref{fig:KL-isocontours} (see, e.g., \citet{Vokrouhlicky98} or
\citet{Subr05} for a general discussion of this topic). Panel E demonstrates that
such a disc is a stronger perturbation than a ring (averaged body) in the sense
that for identical values of parameters ($R_\mathrm{d}$, $a$ and $c$), the topology
is resonant for the disc while it is still non-resonant in the case of the ring
(panel A in Figure~\ref{fig:KL-isocontours}). Another qualitatively new feature
of the isocontours for the case of a disc-like perturbation which appears for
sufficiently low values of $c$ is the existence of
the inner rotational region (panel F).

When the axial symmetry of the perturbing potential is lost by considering the
perturbing body to revolve around the central mass on an eccentric orbit or by
considering an eccentric
stellar disc, the Kozai `integral' $c$ is no longer an integral of motion. Hence,
dimension of the manifold covered densely by the trajectory in the phase space is
higher in comparison to the classical setup and evolution of the test orbit may be
very complex \citep[see, e.g.,][]{Katz11,Lithwick11,Naoz13,Li14,Li14b}. One of the
qualitatively new phenomena are orbital flips, i.e. changes of the sense of the
orbital motion along the
test orbit from prograde to retrograde and vice versa.
During these flips, eccentricity of the test orbit can reach as extremely high values
as $e\sim1-10^{-6}$ \citep{Li14}. Due to non-existence of the Kozai-Lidov integral, it
is also not possible to use the isocontours plotted solely in the $e$-$\omega$ space
of the perturbing potential as a guide through the evolution of the test orbit in contrary to
the case of axially symmetric perturbations. On the other hand, when the rate of change
of $c$ is sufficiently slow, the orbit actually follows the isocontours of the
perturbing potential for a given value of $c$, i.e. it slowly migrates from one
topology to another. As the evolution of $c$ is determined by the octupole term
(and eventually higher order terms) of expansion of the averaged perturbing potential,
we refer to this process as the octupole modulation of the classical Kozai-Lidov
cycles (or simply octupole Kozai-Lidov cycles) further on.
\section{Numerical results}
\label{sec:numerical-results}
Having briefly summarized the key aspects of the Kozai-Lidov dynamics, let us now turn to
identification of its footprints in our numerical models.
\begin{table}
\caption{Variable parameters of the models}
\begin{center}
\begin{tabular}{c|cccc}
 identification & $e_0$ & $\md / \mbh$ & $\mc / \md$ & orientation \\
 \hline
 \model{352} & $0$ & $0.00125$ & $0$ & ---\\
 \model{347} & $0.4$ & $0.00125$& $0$ & random \\
 \model{350} & $0.4$ & $0.00125$& $0$ & aligned \\
 \model{345} & $0.4$ & $0.00125$& $10$ & aligned \\
 \model{344} & $0.4$ & $0.00125$& $100$ & aligned \\
 \model{341} & $\langle 0,\, 0.9\rangle$ & $0.00125$& $0$ & aligned \\
 \model{342} & $\langle 0,\, 0.9\rangle$ & $0.00125$& $10$ & aligned \\
 \model{321} & $\langle 0,\, 0.9\rangle$ & $0.00612$& $0$ & aligned \\
 \model{331} & $\langle 0,\, 0.9\rangle$ & $0.00612$& $4$ & aligned
\end{tabular}
\end{center}
{\bf Description:} $e_0$ is the initial eccentricity of the orbits; in the case
of models \model{341} -- \model{331}, eccentricity is a function
of the semi-major axis according to formula
$e_0 = 0.9 (a - a_\mathrm{min}) / (a_\mathrm{max} - a_\mathrm{min})$; 
$\mbh,\;\md$ and $\mc$ are mass of the central potential, disc and spherical
cluster, respectively; `aligned'
orientation means that all orbits have initially common directions of the apsidal lines,
while `random' corresponds to the case of uniformly distributed longitude of the
ascending node and argument of pericenter.
\label{tab:models}
\end{table}
\subsection{Model and method}
\begin{figure*}
\begin{center}
\includegraphics[width=\columnwidth]{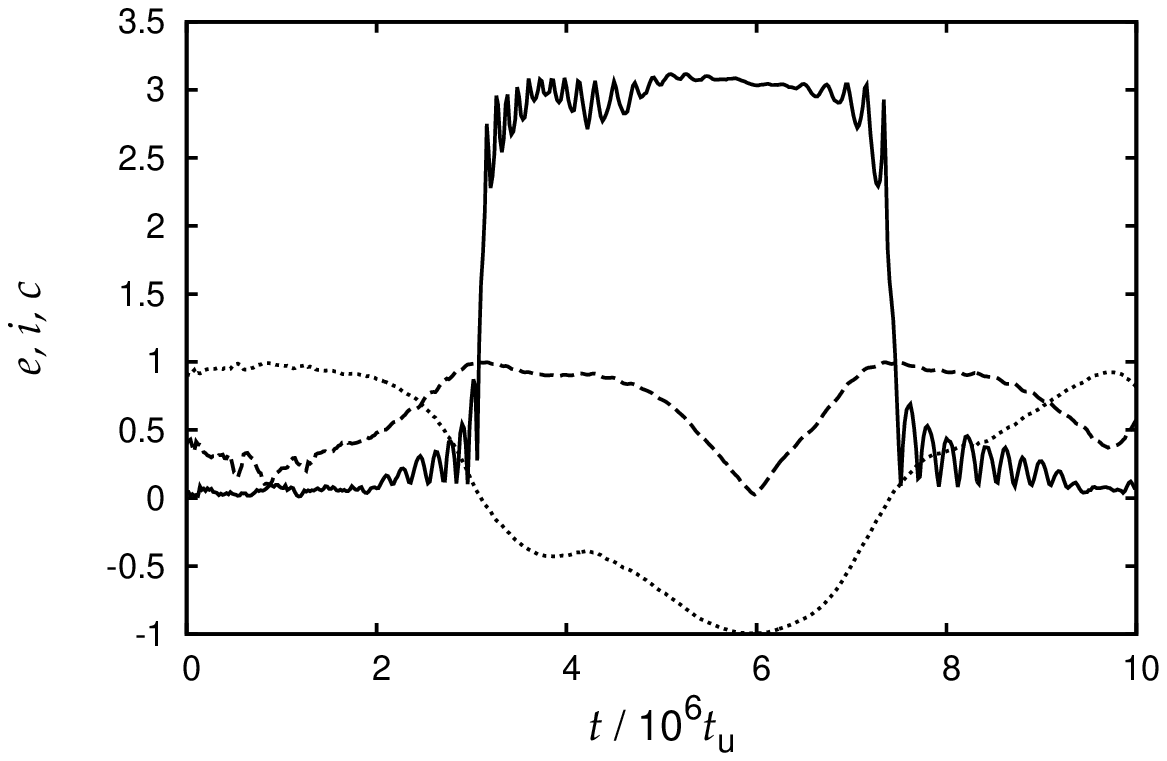}
\hfill
\includegraphics[width=\columnwidth]{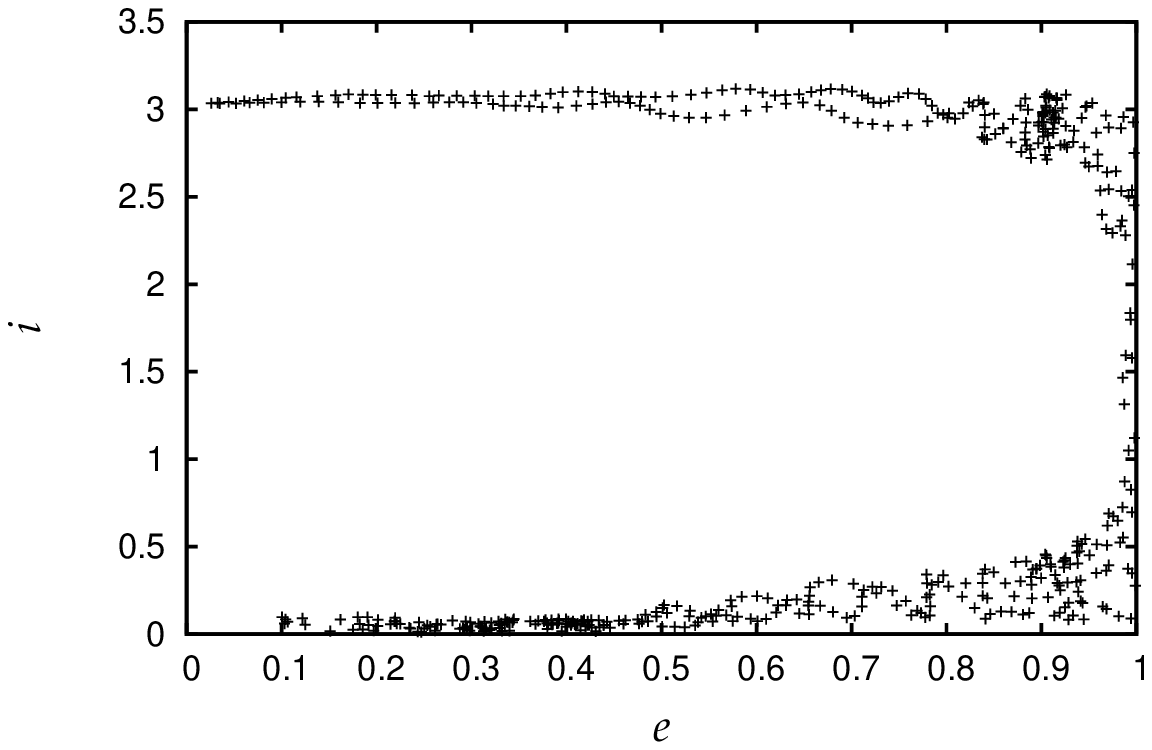}
\caption{\label{fig:coplanar-flip}
A flipping orbit from model \model{350}.
Left panel: Evolution of eccentricity (dashed line), inclination (solid) and the
Kozai-Lidov integral (dotted). Right panel: Evolutionary track of that orbit in
the $e$--$i$ space \citep[cf. right panels of Figures 3 and 4 in][]{Li14}.}
\end{center}
\end{figure*}
\label{sec:models}
First, we introduce several numerical models of a stellar disc around a SMBH which we
used to study the Kozai-Lidov dynamics.
The central SMBH is simulated by a fixed Keplerian potential of point mass $\mbh$
which we also use as a definition of mass unit, i.e. $\mbh = 1\,\unit{M}$. The stellar
disc of total mass $\md$ is represented by $500$ particles of equal mass.
Except for one model, we considered $\md = 0.00125\,\mbh$ (see Table~\ref{tab:models}).
Semi-major axes are
generated according to the distribution function $\dist{a}(a) \propto a^{-1}$ in
$\langle 10\unit{r}, 100\unit{r}\rangle$ which approximately corresponds to a disc
surface density $\propto r^{-2}$. Here, $\unit{r}$ represents an arbitrary length unit
and, together with $\unit{M}$ it can be used for a natural definition of the unit
of time\footnote{For a natural target system, the Galactic Center, our models
scale as: $\unit{M} \approx 4\times10^6\, \msun,\; \unit{r} \approx 0.004\,\pc$
and $\unit{t} \approx 1.9\,\mathrm{yr}$.}
, $\unit{t} \equiv \sqrt{\unit{r}^3 / G\unit{M}}$, where $G$ stands for the
gravitational constant. Initial
inclinations are generated according to the distribution function $\propto \sin i$
in an interval $\langle 0, 2\degm \rangle$, i.e. the normal vectors of the orbital
planes are uniformly distributed within a cone with the half-opening angle $2\degm$.
Values of the initial eccentricities and orientations of the orbits are described in
Table~\ref{tab:models}. Finally, an optional component of the model is a spherically
symmetric gravitational potential $\Phi_\mathrm{c} \propto \sqrt{r}$ which
corresponds to a smooth distribution of mass with density $\varrho_\mathrm{c}(r)
\propto r^{-3/2}$. It stands for a spherically symmetric star cluster parametrized
by its mass, $\mc$, enclosed within the radius $100\,\unit{r}$, i.e. approximately
within the radial domain of the disc.

We concentrated solely on the dynamics of the studied system, i.e. all stars
were treated as point masses. Equations of motion were integrated by means of the
{\nbody} integration code NBODY6 \citep{Aarseth03} which has been modified
by addition of the external potential of the SMBH and the spherical cluster. We have
also added an option for monitoring the minimal value of the radial coordinate
(i.e. distance to the SMBH) for each particle.
\subsection{Kozai-Lidov oscillations}
\label{sec:KL-oscillations}
In our calculations, we have met various modes of the Kozai-Lidov dynamics, depending
on the particular setups of the integrated models. In the following paragraphs, we
demonstrate them on selected individual trajectories. Let us, however, mention that,
due to the stochastic nature of the considered {\nbody} system, the sample orbits lack
the purity of secular evolution often seen within the systems investigated in the
scope of the celestial mechanics. Still, the orbits presented below
belong to those which enable us to identify the particular mode of the Kozai-Lidov
dynamics. Beside them, we have observed many other orbits undergoing some kind of
coupled eccentricity and inclination oscillations which, however, were difficult to
be undoubtedly categorized.

\begin{figure*}
\begin{center}
\includegraphics[width=\columnwidth]{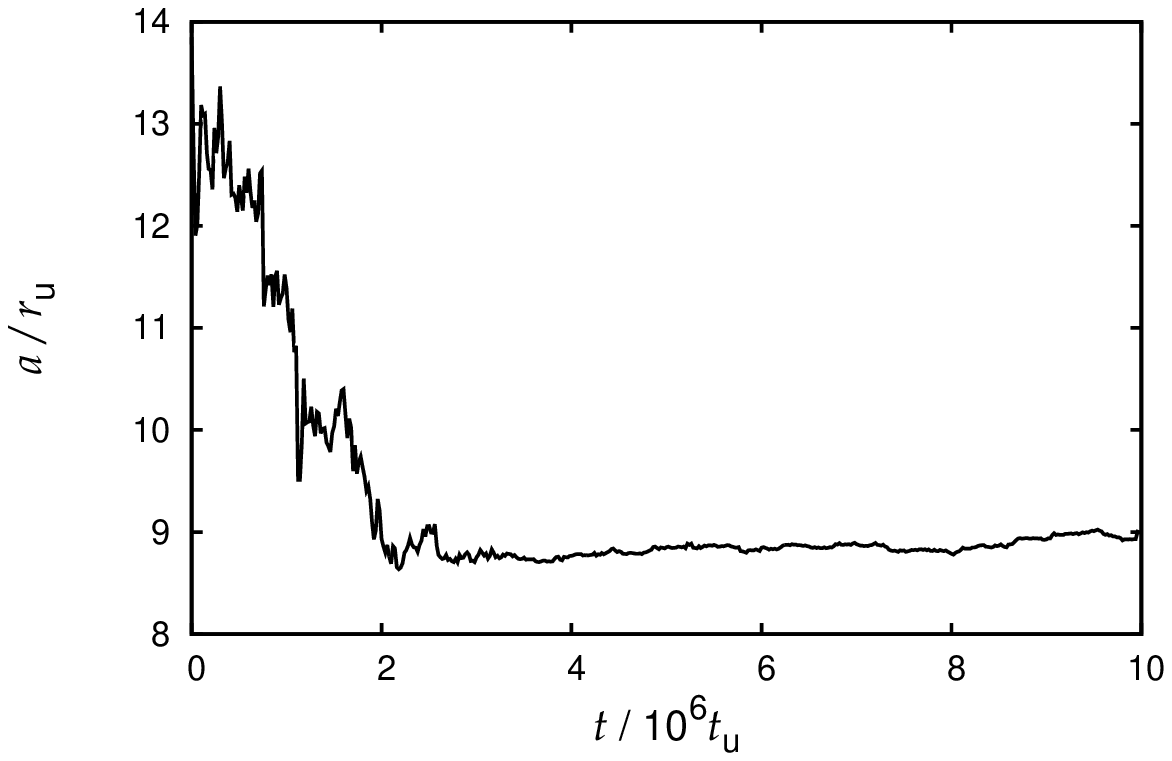}
\hfill
\includegraphics[width=\columnwidth]{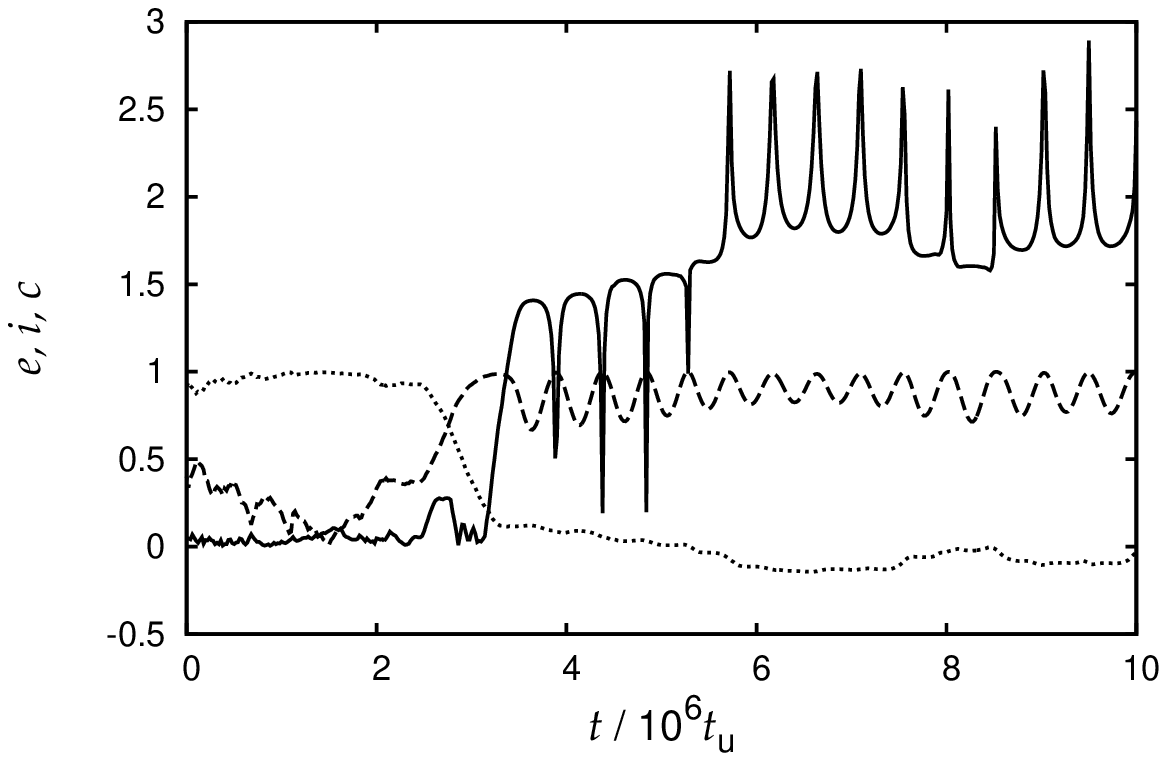}
\caption{\label{fig:star-83}
 An oscillating orbit in model \model{345}. Left: Evolution
 of the semi-major axis. Right: Evolution of the eccentricity (dashed
 line), inclination (solid) and the Kozai-Lidov integral (dotted). }
\end{center}
\end{figure*}
\begin{figure*}
\begin{center}
\includegraphics[width=\columnwidth]{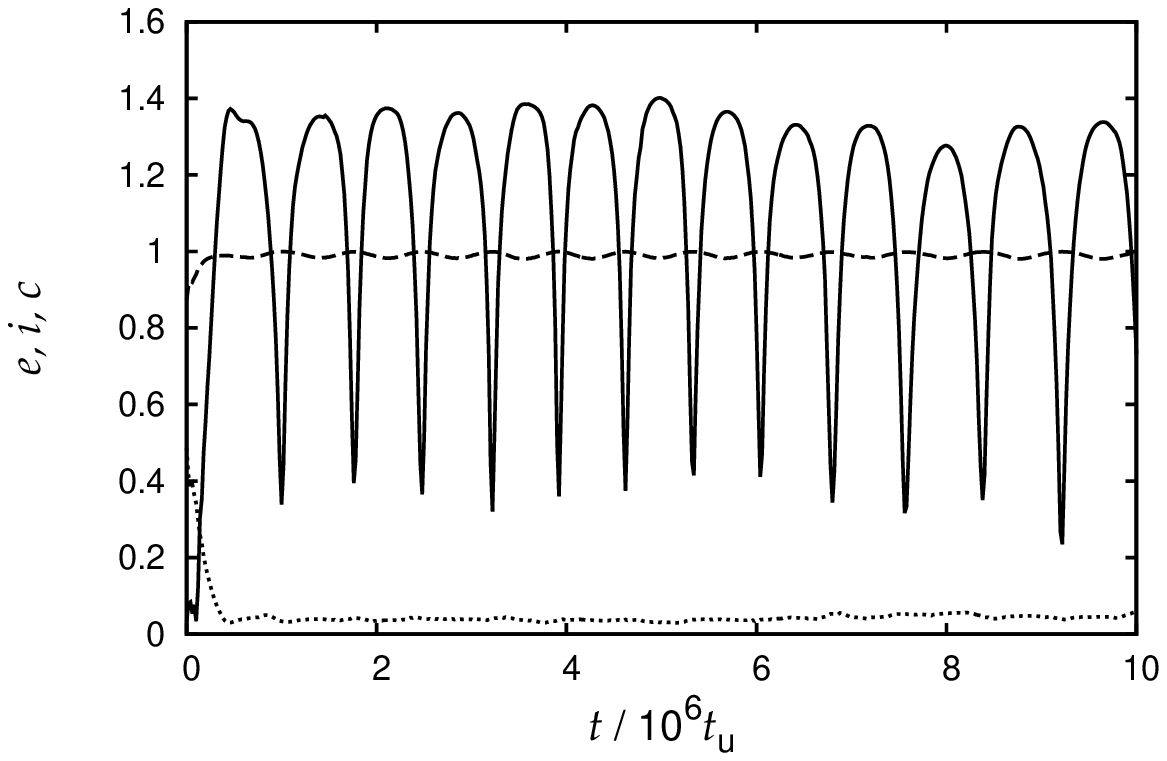}
\hfill
\includegraphics[width=\columnwidth]{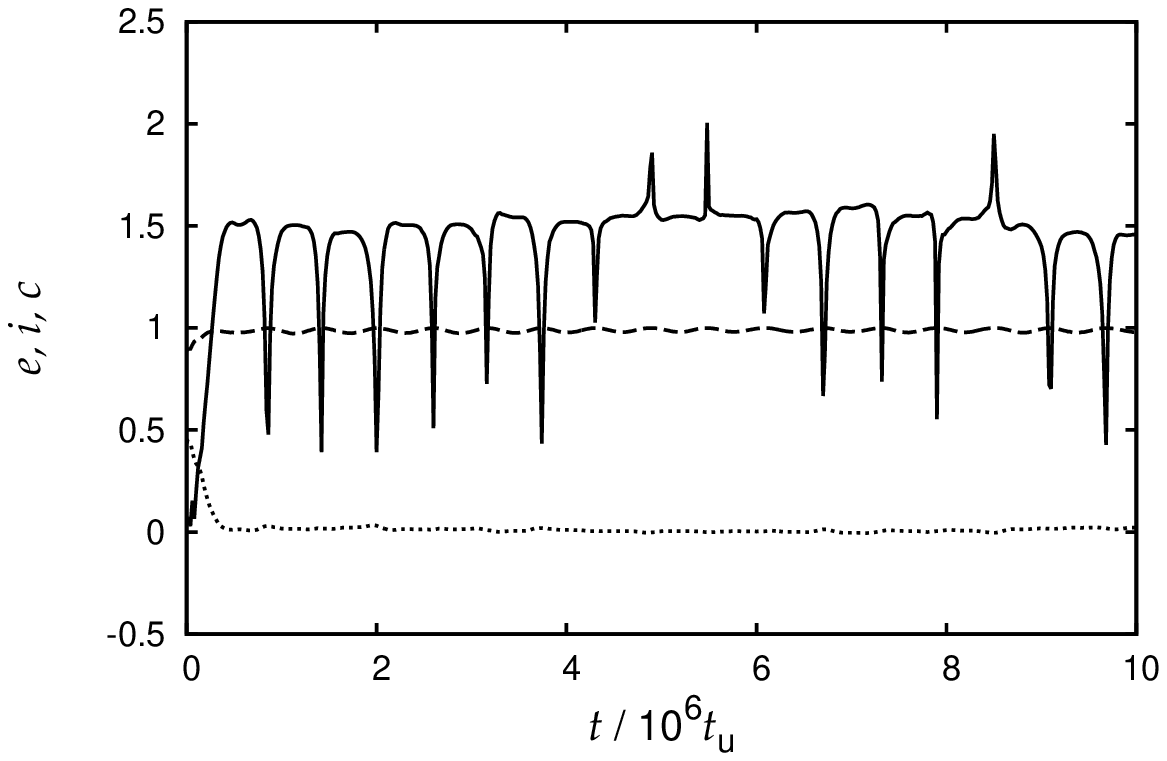}
\caption{\label{fig:outer-Kozai}
Two orbits from model \model{331} undergoing the outer Kozai-Lidov cycles.
Evolution of eccentricity, $e$,
inclination, $i$ and the Kozai-Lidov integral, $c$, is plotted with dashed, solid and
dotted line, respectively \citep[cf. Figure~2 in][]{Bailey92}.}
\end{center}
\end{figure*}
The most prominent effect which is invoked by a non-zero eccentricity of the stellar
disc is the flipping of the orbits in its innermost parts.
This kind of evolution can be best observed in model \model{350} which lacks the
additional spherical potential; an example of a flipping orbit is shown in
Figure~\ref{fig:coplanar-flip}. A rather periodic secular evolution of $c$ (left panel,
dotted line) indicates that the octupole term of the perturbing potential plays an
important role. More specifically, both the slow secular change of
eccentricity (dashed line) as well as the nearly step-wise fashion
change of inclination (solid line) are characteristic for the so-called coplanar
flipping phenomenon \citep{Li14}. An additional piece
of evidence for this particular type of the octupole modulation of the classical Kozai-Lidov cycles is the
shape of the area covered by the evolutionary track of the
flipping orbit in the $e$-$i$ space \citep[right panel; cf.][Figure 4]{Li14}.

Once the disc is embedded in a spherical cluster, its evolution changes.
The coplanar flipping in the innermost parts of the disc becomes less frequent, as
fewer orbits can reach the octupole Kozai-Lidov resonances. Weakening of the systematic
secular evolution leads to promotion of chaotic pattern of orbital trajectories in
the phase space. An example of the better-behaved (i.e. less chaotic) orbits is
displayed in Figure~\ref{fig:star-83}.
We can see that it first evolves rather stochastically in the complex {\nbody}
setting, however, showing a systematic decrease of the semi-major axis $a$ (left
panel). At $t\approx2\times 10^6\,\unit{t}$, the fashion of the evolution changes. Semi-major axis $a$
stops decreasing and remains roughly constant further on. On the other hand,
the value of the Kozai-Lidov integral $c$ (right panel, dotted line) starts to decrease
in a very smooth way from nearly unity to a near-zero value reached at
$t\approx3\times 10^6\,\unit{t}$.
As the time-scale of this decrease covers $\sim10^4$ orbital periods, it appears
that it is of a secular rather than stochastic nature, presumably being a fragment
of an octupole Kozai-Lidov cycle in the overall non-axisymmetric potential of the
disc. When the Kozai-Lidov integral $c$ stops decreasing, it remains roughly constant
for the rest of the investigated time interval, showing only random-walk-fashion
fluctuations. At the same time, mutually coupled periodic oscillations of eccentricity
$e$ and inclination $i$ occur (right panel, dashed and solid line), during which
$e$ reaches nearly unity in its maxima. This indicates that, from
$t\approx3\times10^6\,\unit{t}$ on,
the orbit undergoes the classical Kozai-Lidov cycles in
the global potential of the disc. We can further speculate that their onset occurred
when the (initially very high) value of $c$ became sufficiently low for the Kozai-Lidov
resonance to appear, i.e. for the topology of the perturbing potential isocontours
to change from non-resonant to the resonant one (see Section~\ref{sec:KL}).
The orbital flip that took place at $t\approx5.5\times10^6\,\unit{t}$ is likely to
be a result of a random two-body encounter that changed the sign of the near-zero $c$.
Note that oscillations of individual orbits were also
reported by \citet[][Fig.~1]{Madigan09} in a setup which is similar to our model
\model{345}. We suppose that also in their case, the driving mechanism was the
Kozai-Lidov resonance.

\begin{figure*}
\begin{center}
\includegraphics[width=\columnwidth]{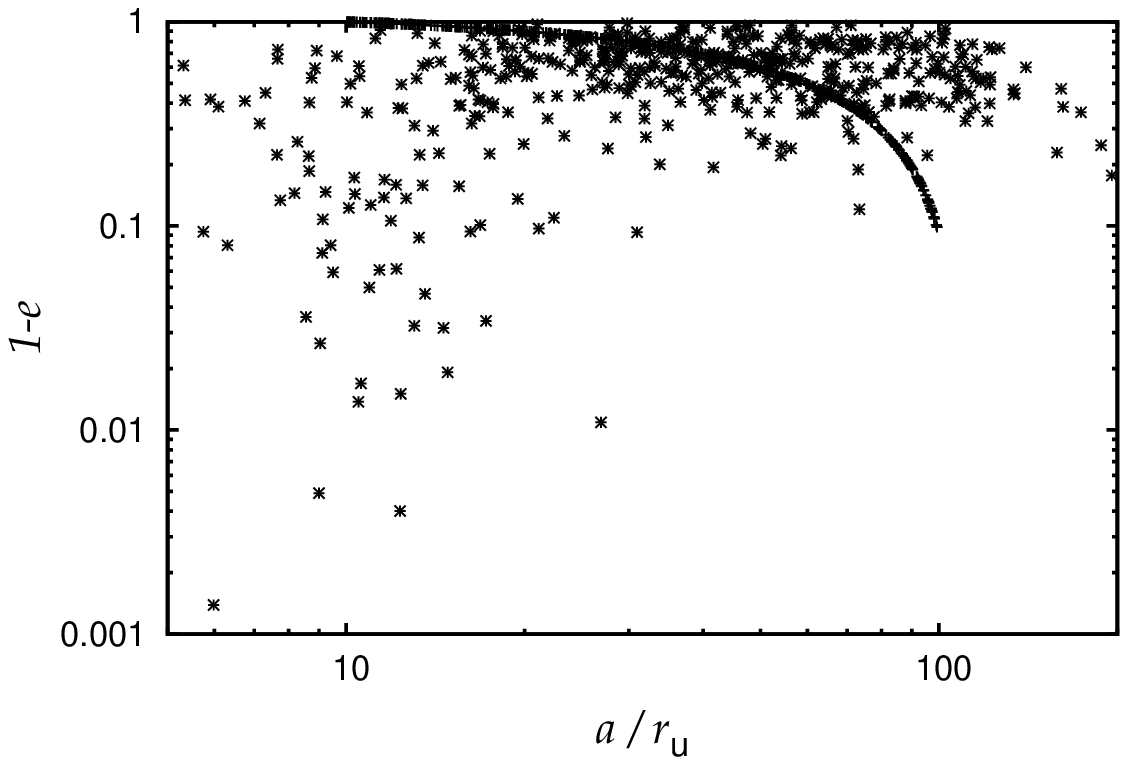}
\hfill
\includegraphics[width=\columnwidth]{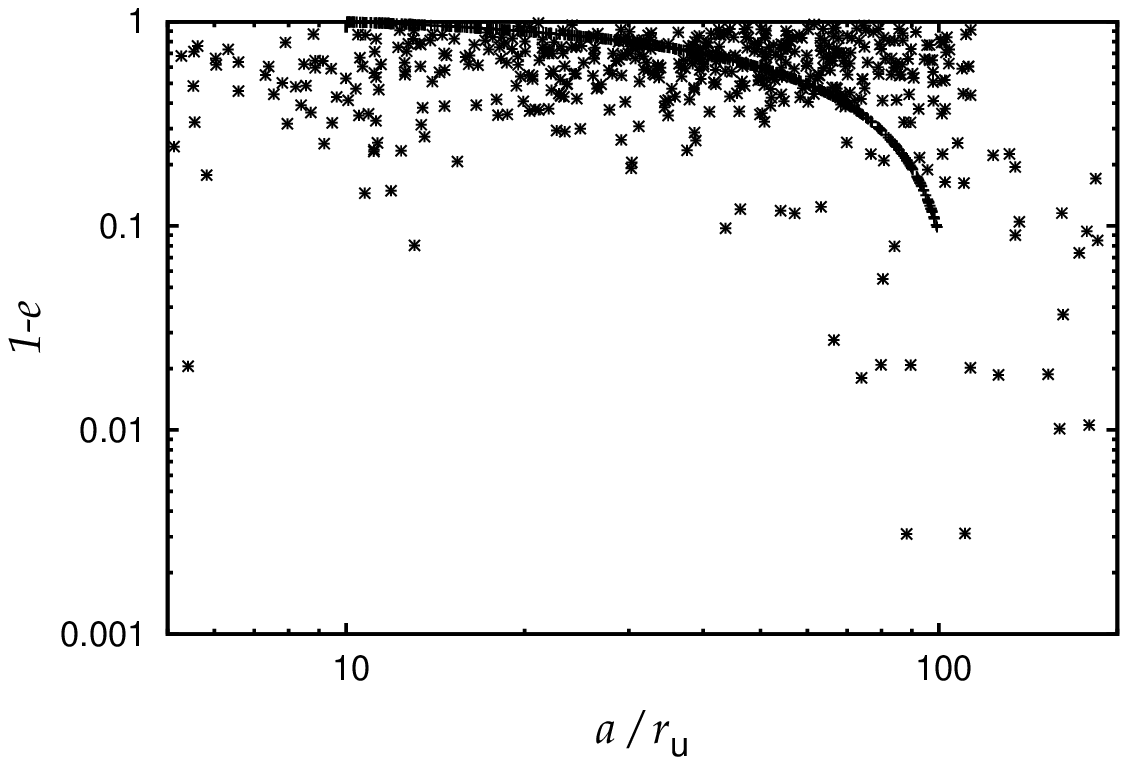}
\caption{\label{fig:Kozai-ae}
Distribution of stellar orbits in the $a$--$e$ space for models \model{321} (left)
and \model{331} (right). Plus signs that form a continuous line represent the initial
state, while the crosses show overplot of three states at $t=10^5\,\unit{t},\;
10^6\,\unit{t}$ and $10^7\,\unit{t}$.}
\end{center}
\end{figure*}

Yet another type of oscillations
occur in the outer parts of the disc when it is embedded in the spherical cluster
(i.e. they were not present in an isolated disc).
The affected orbits undergo the so-called outer Kozai-Lidov
oscillations \citep[e.g.][]{Bailey92,Thomas96, Gallardo12}. Examples of such
oscillating orbits, which are particularly  common in model \model{331}, are shown
in Figure~\ref{fig:outer-Kozai}. The presence of these orbits is, at a first glance,
a rather unexpected result as the spherical potential of the cluster generally damps
such resonant effects. However, in spite of the fact that the potential of the cluster
indeed decreases volume of the resonant area in the phase space and pushes it to
higher eccentricities, the same potential can modify the internal flow of angular
momentum in the disc, increasing its ability to push the stellar orbits to high
eccentricities, i.e. to the resonant area. \NEW{The role of the spherical cluster
in triggering the Kozai-Lidov cycles in the outer parts of the disc is visualized
in Figure~\ref{fig:Kozai-ae} which shows positions of stellar orbits in the $a$--$e$
space for models \model{321} (without the spherical cluster) and \model{331}.
In the former case, we see extreme eccentricities of several orbits from the inner
parts of the disc which undergo the octupole Kozai-Lidov cycles (their pattern is
similar to the example trajectory presented in Figure~\ref{fig:coplanar-flip}).
However, when the spherical stellar cluster is included, the oscillating orbits
at the inner edge of the disc are suppressed and we observe extreme eccentricities
at the outer edge of the disc.}

\subsection{Statistical view on the Kozai-Lidov oscillations}
\begin{figure}
\begin{center}
\includegraphics[width=\columnwidth]{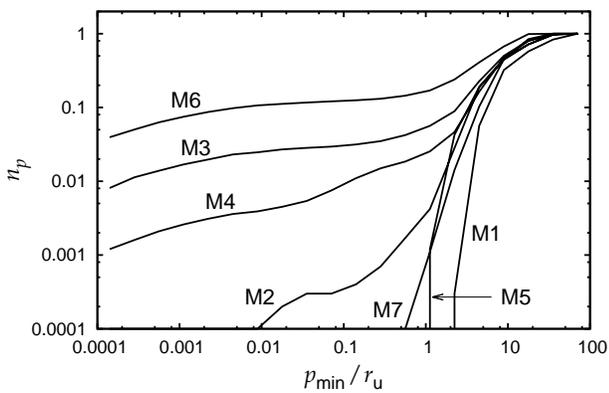}
\caption{\label{fig:pericentres}
Cumulative distribution functions of the closest approaches, $p_\mathrm{min}$, of
the stars to the SMBH in various models (averaged over 20 realizations in all cases).}
\end{center}
\end{figure}
A common feature of all of the above described oscillations are high values of
eccentricity reached during its maxima. One of the methods how to estimate the
effectivity of the individual models in producing high-eccentricity orbits is to
evaluate the closest approaches of the stars to the central SMBH during the
whole integration and construct their cumulative distribution function. The results
for the models introduced in this paper are shown in Figure~\ref{fig:pericentres}.
In agreement with intuition, there are no high-eccentricity orbits in the initially
circular disc (\model{352}). When the orbits are initially eccentric but
randomly oriented (\model{347}), some of them are pushed to high
eccentricities (presumably) due to mutual interactions, however, their number is
still very low. The situation changes when the orbits are initially aligned --
in models \model{350} and \model{341},
a large fraction of orbits reach eccentricities higher than $0.9$. A substantial
fraction of those in the inner parts of the disc undergo the coplanar flipping.

Subsequent models \model{345} and \model{344} share all but
one parameter, $\mc$, with model \model{350} and they demonstrate the damping
effect of the spherical cluster. When $\mc = 10 \md$ (model \model{345}),
a non-negligible number of oscillating orbits still persist but, unlike for model
\model{350}, most of them are of the quadrupole nature (with $c$ conserved). Increasing further the mass
of the spherical cluster (model \model{344}) leads to a complete damping of Kozai-Lidov
oscillations and the distribution of minima of pericenters becomes similar to
models \model{352} and \model{347} with an axially symmetric disc. This result
is in accord with \citet{Gualandris12} who followed evolution of an eccentric disc
embedded in a cluster of mass $\mc\approx140\md$ and
did not observe any periodic oscillations of orbital elements.

Finally, Figure~\ref{fig:pericentres} also demonstrates sensitivity of the angular
momentum transfer processes on the properties of the disc, in particular, the
initial distribution of eccentricities. In the case of a standalone disc (i.e.
not embedded in the spherical cluster), model \model{341} with eccentricities
increasing from zero to $0.9$ outwards produces by a factor of $\approx 5$ more
extremely oscillating orbits than model \model{350} which is initially composed
of equally eccentric orbits ($e_0 = 0.4$). However, adding a spherical potential
of mass $\mc = 10 \md$ leads to a complete damping of orbital oscillations in the
case of disc with gradient distribution of eccentricities in contrary to the model
with initially equal eccentricities, which is able to produce highly eccentric
orbits in spite of the presence of the extended spherical potential
(cf. models \model{342} and \model{345}).

Let us mention that the production rate of high-eccentricity orbits in our
models may be affected by the fact that we emulate the gravity of the spherical cluster
by a smooth analytic potential. \NEW{The orbit presented in Figure~\ref{fig:star-83}
however, indicates that the secular resonances may be able to trap some orbits
even in the disturbing {\nbody} environment. The particular displayed orbit
undergoes systematic change of the Kozai-Lidov integral during the time interval
$2.5\times 10^6\,\unit{t} \lesssim t \lesssim 3\times 10^6\,\unit{t}$. At the same
time, it is mostly embedded in the disc, whose relaxation time is definitely shorter
than the relaxation time of grainy spherical cluster of reasonable parameters.}
Furthermore, it was already suggested by \cite{Lockmann09} that a grainy
potential of the real astrophysical cluster is not as effective in damping the
Kozai-Lidov resonance as the corresponding smooth potential. Hence, our estimates of
the number of produced high-eccentricity orbits may serve as lower limits.

\begin{figure}
\begin{center}
\includegraphics[width=\columnwidth]{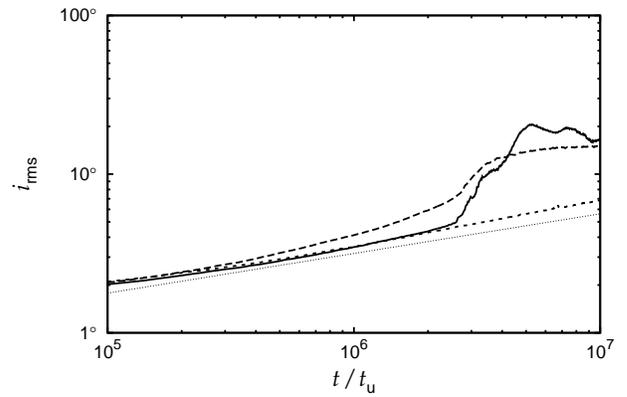}
\caption{\label{fig:irms}
 Evolution of the root-mean-square inclination in the disc for models \model{350},
 \model{345} and \model{347} (solid, dashed and dotted lines, respectively; averaged
 over 20 realizations). Thin dotted line represents slope $t^{1/4}$.}
\end{center}
\end{figure}
Another quantity which reflects the Kozai-Lidov oscillations of individual orbits
from the disc is a root-mean-square value of inclination, $\irms$. In the case of an
initially circular disc, it has been predicted theoretically and also verified by
means of numerical models \citep[e.g.][]{Stewart00} that $\irms$ should grow in time
approximately as $\irms \propto t^{1/4}$ due to the two-body relaxation. As we show
in Figure~\ref{fig:irms} (dotted line), also model
\model{347} which represents an axially symmetric disc formed by randomly oriented
eccentric orbits exhibits the same rate of evolution of $\irms$. A qualitatively
different evolution is observed in the case of model \model{350} (solid line) which
is formed by aligned eccentric
orbits -- from $t\approx 3\times 10^6\, \unit{t}$ on, we see an accelerated growth of
$\irms$. Let us note that such an accelerated growth of $\irms$ has been reported
already by \citet[][Figure 3]{Cuadra08} who integrated a model similar to our one.
Unlike these authors, who attributed this effect to resonant relaxation, we
interpret this behavior as a manifestation of the Kozai-Lidov oscillations.
In particular, in this case, when the disc is not embedded in the additional spherical
potential, we distinguish majority of the oscillations to correspond to the octupole Kozai-Lidov cycles,
often leading to the coplanar flipping during which inclinations of the orbits reach values very close
to $\pi$. The accelerated growth of $\irms$ is then a consequence of averaging over
a sample of orbits which includes a certain number of those with large inclinations.
Let us, however, point out that this growth of $\irms$ does not necessarily
lead to geometrical thickening of the disc. The orbits undergoing octupole Kozai-Lidov
cycles reach extreme eccentricities at the stages of high $\sin i$, i.e. although
being highly inclined, they are still embedded in the thin disc structure. 

The picture, however, changes when the disc is embedded in the spherical cluster
(model \model{345}; dashed line in Figure~\ref{fig:irms}). In this case, the coplanar
flips are substantially damped, which leads to a less prominent jump of $\irms$ at
$t\approx 3\times10^6 \unit{t}$. On the other hand, evolution of $\irms$ within
model \model{345} deflects from the $t^{1/4}$ line already at $t < 10^6\, \unit{t}$,
which we found to be due to presence of the quadrupole Kozai-Lidov oscillations with
conserved $c$. This, in contrary to the octupole modulated case, means that the
oscillating orbits reach minima of their eccentricities at the high inclination stage,
i.e. they spend non-negligible period of time above the equatorial plane and
the value of $\irms$ reasonably well represents the geometrical thickness of the disc.
\section{Conclusions}
We analyzed the orbital evolution of an initially thin and eccentric stellar disc
around the central supermassive black hole (SMBH). When the stellar orbits are
initially randomly oriented, the overall evolution is similar to that of a circular
disc. If the orbits share a common orientation of their apses, the evolution of the
disc is significantly different, comprising a variety of the Kozai-Lidov oscillations of
eccentricity and inclination of different types. These strongly
depend on the particular initial setup; among all the possible variations, we have
explicitly discussed the role of an embedding spherical cluster and the eccentricity
distribution within the disc.

The onset of the Kozai-Lidov oscillations appears to be a rather generic process
occurring in eccentric stellar discs. Its systematic nature can successfully compete
with the chaotic two-body relaxation, preferably in the less dense parts of the disc.
In our particular setup, we have found most of the oscillating orbits to appear in
the outer part of the disc and in the innermost one, below the initial inner radius.
The Kozai-Lidov resonance is substantially affected by the presence of a global
spherical potential (e.g. of an embedding cluster). In addition to its well-known
damping influence on the Kozai-Lidov oscillations, however, we found that the presence
of a moderately strong spherical potential may actually trigger these oscillations in
the outer parts of the disc due to affecting the internal flow of angular momentum in the
disc. The type of the Kozai-Lidov oscillations strongly depends on the particular setup.
The same holds for the number of oscillating orbits which varies between the order of
$\sim10\%$ in the most favorable case of an isolated disc and $\lesssim1\%$ for a disc
embedded in a spherical potential (see Figure~\ref{fig:pericentres}).

Direct consequence of the numerous oscillating orbits is modification of initial
distributions of their orbital elements. From the observational point of view, the
most interesting is the accelerated growth of root-mean-square inclinations with respect
to the two-body relaxation driven evolution. We have found that in the case of an isolated
disc, this process is the most prominent, being caused by the coplanar flipping.
However, in this case, the highly inclined orbits lie geometrically in the thin disc
structure. On the other hand, well above the original plane of the disc (up to $\pi/2$)
are pushed the
oscillating orbits found in the models with intermediate mass of the embedding spherical
cluster. Another possibly observable feature is the presence of highly eccentric orbits
in the maxima of their Kozai-Lidov cycles.

The Kozai-Lidov oscillations themselves are difficult to be directly identified in
observed astrophysical systems
-- resolution of our telescopes is not sufficient to determine the orbital elements
of stars in foreign galactic nuclei. Even in the case of the Galactic Center,
where the orbital elements of several stars have already been determined, the
time-scale of the Kozai-Lidov oscillations exceeds the human life-time by several
orders of magnitude. One possible indirect evidence of the Kozai-Lidov oscillations
of stellar orbits in galactic nuclei (even the distant ones) are tidal disruptions
of stars which are supposed to occur whenever the eccentricity and, consequently,
the distance to the SMBH during the pericenter passage reaches
a critical value. Being scaled to the nuclear stellar system of the Milky Way,
the tidal disruption radius of a Solar-type star is of the order of
$10^{-6}\,\pc \approx 2.5\times10^{-4}\,\unit{r}$, i.e. in the most favorable case
of a disc not embedded in a spherical cluster, we predict somewhat more than one per
cent of stars from the disc to undergo the tidal disruption event (see models
\model{350} and \model{341} in Figure~\ref{fig:pericentres}). Even though this is
a relatively large number, it is also likely to
expect numerous tidal disruptions of stars from the embedding spherical cluster
due to the Kozai-Lidov cycles \citep{Karas07}. Origin
of a particular tidally disrupted star would thus be uncertain. Moreover, in contrary
to \cite{Karas07}, in our current analysis, we have omitted general relativistic
effects, such as the pericenter shift which is known to suppress the Kozai-Lidov
oscillations \citep{Blaes02}. In the context of the stellar disc around the SMBH,
orbits reaching the stellar tidal disruption radii would suffer from this damping.

Within the context of our Galaxy, even more exciting appears to be tidal break-ups
of binary stars which would occur at radii $\lesssim 10^{-3}\pc$, depending on their
intrinsic orbital parameters. \NEW{Tidal break-ups of binaries have already
been suggested by \cite{Hills88} as a process that may bring stars to orbits tightly
bound to the SMBH and, at the same time, accelerate their former binary companions
to velocities greatly exceeding escape velocity from the Galaxy.} Later on, both
groups of stars were indeed found---the so-called S-stars \citep[e.g.][]{Ghez05}
which are young
stars orbiting the central SMBH with the semi-major axes $\lesssim0.04\,\pc$ and
the hyper-velocity stars which are young stars escaping from the Milky Way with
velocities exceeding $500\,\kms$ \citep[e.g.][]{Brown14}.
Here we suggest, and in a subsequent paper
\citep{Subr15} further investigate, the possibility that all of them may have been born
in a thin eccentric stellar disc and pushed to the tidal radius via the Kozai-Lidov
oscillations.
\section*{Acknowledgments}
We thank David Vokrouhlick\'{y} and the anonymous referee for helpful comments
on the manuscript. We also
acknowledge support of the Czech Science Foundation through
the Project of Excellence No 14-37086G.

\end{document}